\documentclass[twocolumn, amsmath,amssymb,prb]{revtex4}
\usepackage{graphicx,dcolumn,bm,amsmath,amssymb}
\usepackage{pslatex}

\begin{document}
\title{Charge density wave fluctuations, heavy electrons, and superconductivity in   KNi$_2$S$_2$}
\author{James R. Neilson}
\email{jneilso2@jhu.edu}
\affiliation{Department of Chemistry,  Johns Hopkins University, Baltimore, MD 21218}
\affiliation{Institute for Quantum Matter, and Department of Physics and Astronomy,  Johns Hopkins University, Baltimore, MD 21218}

\author{Anna Llobet}
\affiliation{Lujan Neutron Scattering Center, Los Alamos National Laboratory, MS H805, Los Alamos, NM 87545}

\author{Jiajia Wen}
\affiliation{Institute for Quantum Matter, and Department of Physics and Astronomy,  Johns Hopkins University, Baltimore, MD 21218}

\author{Matthew R. Suchomel}
\affiliation{Advanced Photon Source, Argonne National Laboratory, Argonne, Illinois 60439}

\author{Tyrel M. McQueen}
\email{mcqueen@jhu.edu}
\affiliation{Department of Chemistry,  Johns Hopkins University, Baltimore, MD 21218}
\affiliation{Institute for Quantum Matter, and Department of Physics and Astronomy,  Johns Hopkins University, Baltimore, MD 21218}

\begin{abstract}
Understanding the complexities of electronic and magnetic ground states in solids is one of the main goals of solid-state physics.  Materials with the canonical ThCr$_2$Si$_2$-type structure have proved particularly fruitful in this regards, as they exhibit a wide range of technologically advantageous physical properties described by ``many-body physics,'' including high-temperature superconductivity and heavy fermion behavior.  
Here, using high-resolution synchrotron X-ray diffraction and time-of-flight neutron scattering, we show that the isostructural mixed valence compound, KNi$_2$S$_2$, displays a number of highly unusual structural transitions, most notably the presence of charge density wave fluctuations that \emph{disappear} on cooling.
This behavior occurs without magnetic or charge order, in contrast to expectations based on all other known materials. Furthermore, the low-temperature electronic state of KNi$_2$S$_2$ is found to exhibit many characteristics of heavy-fermion behavior, including a heavy electron state ($m^*/m_e \sim$  24), with a negative coefficient of thermal expansion, and superconductivity below $T_c$ = 0.46(2)~K.  In the potassium nickel sulfide, these behaviors arise in the absence of localized magnetism, and instead appear to originate in proximity to charge order.  
\end{abstract}
\maketitle


\section{Introduction}
Quantum coherence of electronic states in metals,  or more generally ``many-body,'' emergent phenomena, such as superconductivity, result from electron-electron or electron-phonon interactions established by the constraint of the lattice.  
Many materials that give rise to many-body physics (\emph{i.e.}, high-temperature superconductivity or heavy-fermion behavior) are comprised of layers of edge-sharing $[MX_4]$ tetrahedra, where $M$ is a transition metal and $X$ is a main-group element, as commonly found in the ThCr$_2$Si$_2$  (\emph{e.g.}, Ba$_{1-x}$K$_x$Fe$_2$As$_2$,\cite{Rotter:2008p5450}  K$_x$Fe$_{2-y}$Se$_2$,\cite{Sun:2012jl} or URu$_2$Si$_2$ \cite{Palstra:1985p7047})  or ZrCuSiAs  (\emph{e.g.}, SmFeAsO$_{1-x}$F$_x$ \cite{Chen:2008fk}) structure-types.  
Extensive work has found rich electronic phenomena in these materials, including hidden-order in URu$_2$Si$_2$,\cite{Tripathi:2007fk} nematic order, \cite{Chuang08012010, Chu13082010} valley density wave order,\cite{PhysRevB.80.024512} magnetoelastic coupling,\cite{PhysRevB.82.020408, Caron_2011,Caron_2012} and Fermi surface nesting.\cite{Cvetkovic:2009p1423}   It is a generally accepted fact that the presence of magnetism and/or magnetic fluctuations are important in producing the correlated electronic behavior in these materials.  

Here, we report that KNi$_2$S$_2$ has a similarly rich structural and electronic phase diagram in the absence of localized magnetism, with several features unexpected under traditional theories of strong electron interactions including (1) the disappearance of charge density wave (CDW) fluctuations concomitant with an unusual \emph{increase} in local symmetry on \emph{cooling} without trivial charge order, and (2) an enhancement of the effective conduction electron mass at low temperatures coupled with negative thermal expansion.  
The former is unexpected on thermodynamic grounds as the increase in local symmetry implies a decrease in the configurational entropy of the structure.  The latter, an increase in electronic entropy, is a hallmark of the many-body ``heavy-fermion'' state, but is unexpected as KNi$_2$S$_2$  shows no signs of the localized magnetism associated with producing such a state.\cite{Coleman:2007ua}

Instead, our findings are most consistent with KNi$_2$S$_2$ harboring electronically driven phase transitions that arise from changes in hybridization of a bath of delocalized conduction electrons with localized and bonded (\emph{i.e.}, CDW) electrons making it an ideal compound for study of the coupling between charge and structural degrees of freedom in mixed-valence materials.  Furthermore, these results demonstrate that proximity to charge order alone, without localized magnetism, can drive strongly correlated physics and warrants further experimental and theoretical attention.

\section{Methods}

Polycrystalline, lustrous and orange-yellow powder of KNi$_2$S$_2$ was prepared as previously described, but with a substitution of S for Se.\cite{Neilson_JACS_2012}  All samples were prepared and handled exclusively inside an argon-filled glovebox; no impurities were detected by laboratory X-ray diffraction.  High-resolution synchrotron X-ray diffraction data were collected using the high-resolution powder diffractometer at the Advanced Photon Source on beamline 11-BM \cite{wang:085105} from polycrystalline powders sealed in an evacuated fused silica capillary backfilled with $p_\text{He}$ = 10 torr.  Data for $T\le$ 100~K were collected using a He cryostat (Oxford Instruments); data collected at $T \ge $ 100~K were collected using a nitrogen cryostream.  Rietveld analyses were performed using GSAS/EXPGUI.\cite{gsas,expgui}  The synchrotron X-ray diffraction (SXRD) data revealed the presence of 3~wt\% K$_2$Ni$_3$S$_4$ impurity and a 1~wt\% Ni$_3$S$_2$ impurity, which were included in the Rietveld analyses.  The chemical occupancies were refined in Rietveld analysis of the SXRD data to test if the KNi$_2$S$_2$ phase was substoichiometric; the values varied less than 1\% from 1, thus the values were fixed to unity.

Neutron total scattering data were collected at temperatures between 5~K and 300~K on polycrystalline KNi$_2$S$_2$ (loaded in a vanadium can with a He atmosphere) using the time-of flight HIPD and NPDF instruments at the Lujan Center, Los Alamos Neutron Science Center, Los Alamos National Laboratory.  PDF analysis was performed on the total neutron scattering data and $G(r)$ were extracted with $Q_\text{max} = 29$ \AA$^{-1}$ (HIPD) and  $Q_\text{max} = 35$ \AA$^{-1}$ (NPDF) using PDFgetN.\cite{pdfgetn}    Least-squares fits to the PDF were performed using PDFgui.\cite{pdfgui}   Reverse Monte Carlo simulations of 20$\times$20$\times$6 supercells (24000 atoms, $\sim$75 \AA/side) were performed using RMCprofile,\cite{Tucker:2007vt} while applying a small penalty for breaking tetrahedral coordination and a hard-sphere cut off for the Ni--S bond distance. The structural parameters (Ni--Ni distance and bond valence sums) were compiled from the average of four independent simulations.  Atomistic visualization was accomplished using VESTA.\cite{vesta}

Physical properties were measured using a Physical Properties Measurement System, Quantum Design, Inc; for measurement below 1.8 K, a dilution refrigerator option was used.  Specific heat measurements were performed using the quasi-adiabatic heat-pulse technique on sintered polycrystalline pellets attached to the sample stage using thermal grease.  Magnetization measurements were carried out at $\mu_0H =$ 1 T and 2 T, and with the susceptibility estimated as $\chi \approx \Delta M/\Delta H = [M_{\text{2T}}-M_{\text{1T}}]/[1\text{T}]$.  Isothermal, field-dependent magnetization measurements were performed over a range of temperatures and at fields from $\mu_0H =$ 0 to 9 T to determine the fraction of impurity spins that contribute to $\chi$ at low-temperatures.  A self-consistent global fit of the field-dependent magnetization data sets was performed to a Brillouin function for impurity paramagnetic spins (with a single set of three parameters $g=2$, $J$, and concentration, fixed to be the same at all temperatures) and to a linear function (the true temperature dependent susceptibility).  For resistivity measurements, platinum wires were attached to sintered polycrystalline pellets using silver paste and dried under argon in a four point configuration.  Equivalent results were achieved by using  Ga$_{0.85}$In$_{0.15}$ as a molten solder.

\section{Results}

\begin{figure}[t]
\centering
\includegraphics[width=2.5in]{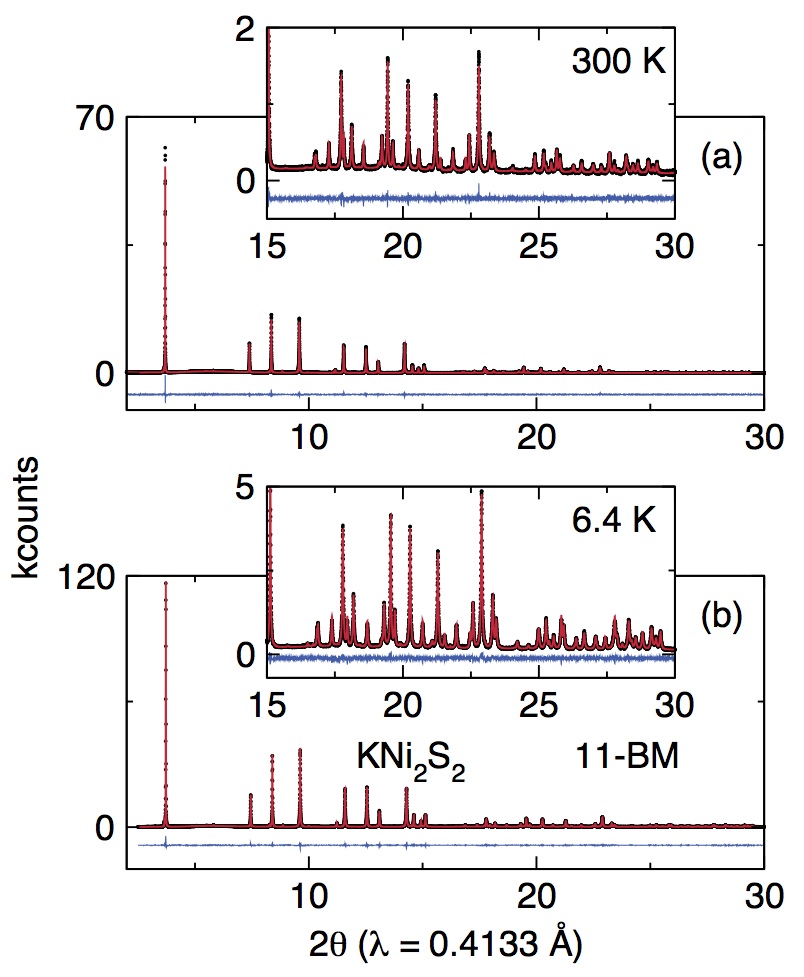}
\caption{Representative Rietveld analyses of SXRD data collected at (a) $T$ = 300~K and (b) $T$ = 6.4~K .  \label{fig:11bmdata}}
\end{figure}

\subsection{Synchrotron X-ray and Time-of-Flight Neutron Scattering}

\begin{figure}[t]
\centering
\includegraphics[width=3in]{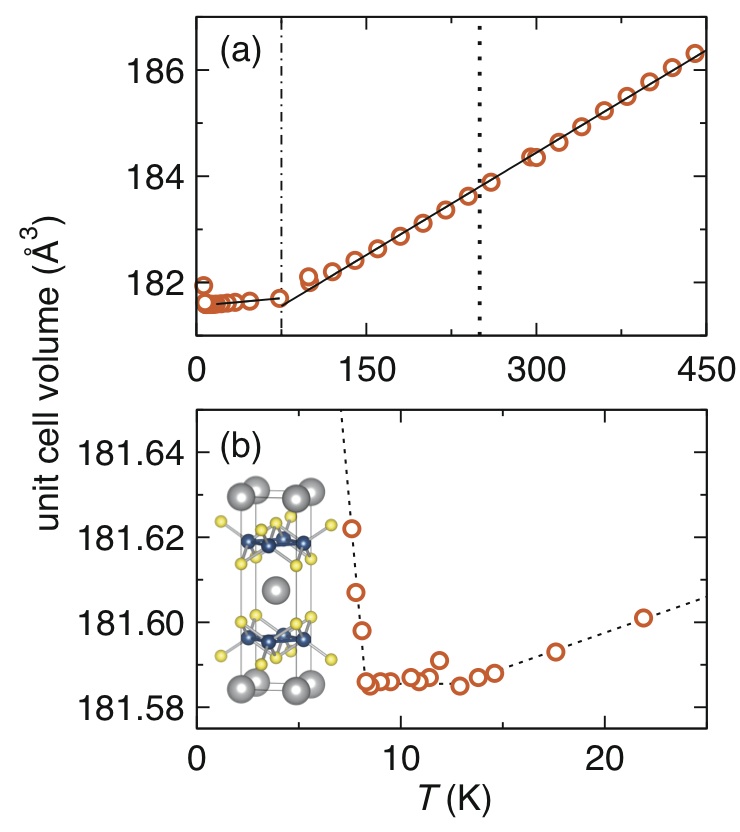}
\caption{(a) From analysis of temperature-dependent high-resolution synchrotron X-ray diffraction (11-BM, Advanced Photon Source), a clear transition in unit cell volume of KNi$_2$S$_2$ is observed at $T\sim$ 75~K (dot-dashed line); no transition is observed at $T=250$ K (dotted line). (b) The temperature dependence of the unit cell volume of KNi$_2$S$_2$ shows two anomalies  below $T<13$~K, with an overall negative coefficient of thermal expansion below $T<8.9$~K.   (dashed lines guide the eye; inset: unit cell of KNi$_2$S$_2$). \label{fig:cellvolume}}
\end{figure}

Analysis of high-resolution synchrotron X-ray diffraction (SXRD) data indicates that the average crystallographic symmetry of KNi$_2$S$_2$ is tetragonal ($I4/mmm$) at all temperatures measured (6.4~K $<T<$ 440~K; Figure~\ref{fig:11bmdata}).  
In KNi$_2$S$_2$, the [Ni$_2$S$_2$]$^-$ layers of edge-sharing [NiS$_4$] tetrahedra are separated by K$^+$ ions [Figure~\ref{fig:cellvolume}]; this leaves the nickel atoms with a formal valence of ``Ni$^{1.5+}$''.    No periodic distortions are found, in contrast to the commensurate distortions observed in KCu$_2$Se$_2$ \cite{PhysRevB.67.134105} or the incommensurate modulation of $\beta$-SrRh$_2$As$_2$.\cite{PhysRevB.85.014109}

\begin{table}
\begin{ruledtabular}
  \caption{Structure parameters of KNi$_2$S$_2$ obtained from Rietveld refinement of data collected  from synchrotron X-ray powder diffraction, showing fractional coordinates ($x,y,z$), chemical site occupancies (occ.), and isotropic atomic displacement parameters ($U_{iso}$) described with the $I4/mmm$ (139) spacegroup.  }
  \label{tab:structure}
    \begin{tabular}{c|c|c|c|c|c|c}
    $T=300$~K  &  \multicolumn{6}{c}{$a$ = 3.79221(6) \AA; $c$ = 12.8193(2) \AA}\\ 
    \hline
    Atom & Site & $x$ & $y$ & $z$ & occ. & $U_{iso}$ (\AA$^2$)\\
    \hline
   K & $2a$ & 0 & 0 &0 & 1 & 0.0199(2) \\
   Ni & $8g$ & 0 & 0.5 & 0.258(1)  & 0.5  & 0.0124(1) \\
      S & $4e$ & 0 & 0  & 0.3500(4) & 1 & 0.0149(1)\\
\hline
\hline
    $T=10.5$~K & \multicolumn{6}{c}{$a$ = 3.7792(1) \AA; $c$ = 12.7139(1) \AA}\\
    \hline
    Atom & Site & $x$ & $y$ & $z$ & occ. & $U_{iso}$ (\AA$^2$)\\
    \hline
   K & $2a$ & 0 & 0 &0 & 1 & 0.0070(3) \\
   Ni & $4d$ & 0 & 0.5 & 0.25  & 1  & 0.0049(1) \\
      S & $4e$ & 0 & 0  & 0.3500(1) & 1 & 0.0075(2)\\
  \end{tabular}\\
\end{ruledtabular}
\end{table}

Nonetheless, detailed analysis of the diffraction data reveals a  structural transition near $T\sim$ 75~K and negative thermal expansion below $T \sim$ 9 K.  The temperature dependence of the unit cell volume, extracted from Rietveld analysis, is shown in Figure \ref{fig:cellvolume}. A 1 wt\% impurity (Ni$_3$S$_2$, undetectable by laboratory XRD) included in the SXRD data refinements acts as an internal standard.  First, there is clear change in slope near $T$ = 75 K. Second, the unit cell volume remains constant from $T$ = 13 to 8.9 K, but then increases with further cooling [Figure \ref{fig:cellvolume}(b)]. The former is consistent with a structural change at $T$ = 75 K,\cite{Megaw} while the latter indicates a switch to negative thermal expansion behavior.

\begin{figure}[t!]
\centering
\includegraphics[width=2.5in]{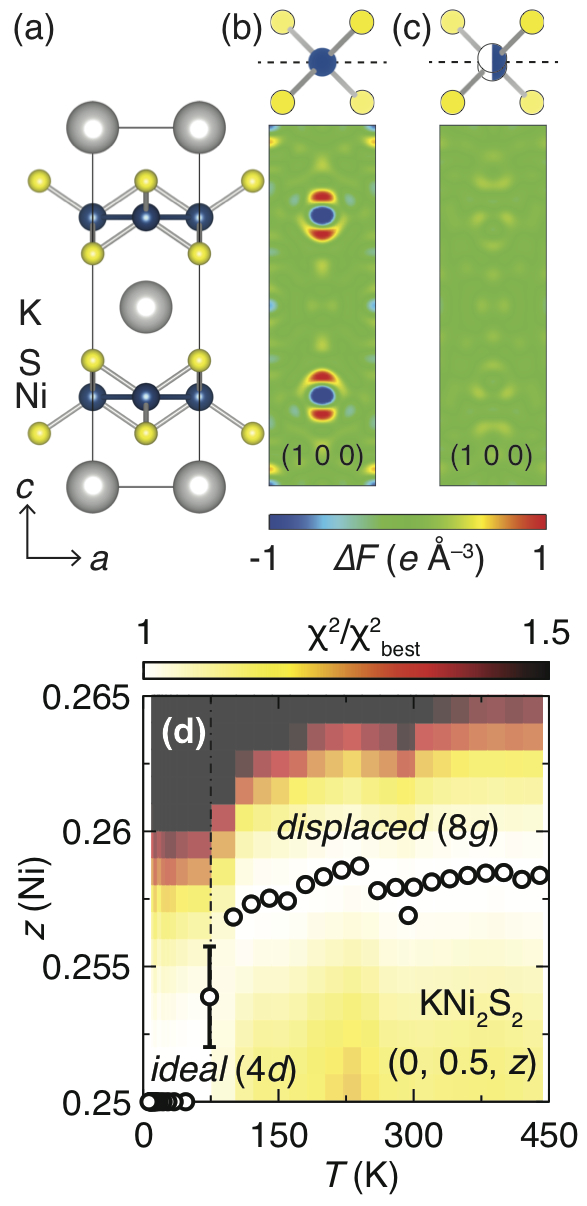}
\caption{
(a) Illustration of the (100) crystal plane of KNi$_2$Se$_2$ and Fourier difference maps from data collected at $T$ = 300~K, viewed on the (001) plane reveal the displacement of Ni atoms from the 
(b) ideal $4d$ (0,~0.5,~0.25) position to the (c) displaced site at $8g$ (0,~0.5,~$z$).  The schematic unit cell on the right illustrates the location of the Ni atoms (blue spheres).  
(d)  Coincident with the transition in unit cell volume are charge density wave (CDW) fluctuations that disappears on cooling as $T<$ 75~K; these CDW fluctuations manifest as a displacement of the Ni atom ($z$) from the ideal $4d$ (0,~0.5,~0.25) site to the $8g$ (0,~0.5,~$z$) site.  Through all measured temperatures, the space-group symmetry, $I4/mmm$ is retained, with the displaced $z$ position requiring a split-site occupancy ($4d$ to $8g$).  The color denotes the normalized Rietveld goodness-of-fit ($\chi^2/\chi^2_\text{best}$; brighter = better fit) to show the significance of the displacement shift at $T\sim$ 75~K. \label{fig:11bmdistortion}    }
\end{figure}

The change in slope in unit cell volume at $T$ = 75 K is coupled to a striking change in the average structural model required to describe the data.  Adequate fits to the SXRD data collected at room temperature are obtained only when the Ni atoms are displaced off of the high symmetry position and are instead statistically distributed in a split-site model with the lower $2mm$ site symmetry on the $8g$ (0,~0.5,~$z$) Wycoff position [Figure~\ref{fig:11bmdata}(a), Table~\ref{tab:structure}].  The strongest evidence for the presence of this distortion comes directly from a Fourier difference map generated from the undistorted structural model, illustrated in Figure~\ref{fig:11bmdistortion}.  A dearth of scattering intensity is located above and below the $4d$ Ni site, while an excess of intensity is located at the Ni position of (0,~0.5,~0.25), illustrated in Figure~\ref{fig:11bmdistortion}(b).  Introduction of the distortion produces an undisturbed Fourier difference map [Figure~\ref{fig:11bmdistortion}(c)].  In contrast, at $T=6.4$~K, the SXRD data are described by an ideal ThCr$_2$Si$_2$ structure, with Ni atoms on the $4d$ (0,~0.5,~0.25) Wycoff position with $\bar{4}m2$ site symmetry [Figure~\ref{fig:11bmdata}(b), Table~\ref{tab:structure}].  The temperature dependence of the off-centering is shown in Figure~\ref{fig:11bmdistortion}(d). There is only a small variation with temperature above  $T >$ 75 K. However, the off-centering abruptly disappears on cooling below $T=$ 75 K.

\begin{figure}[t!]
\centering
\includegraphics[width=3in]{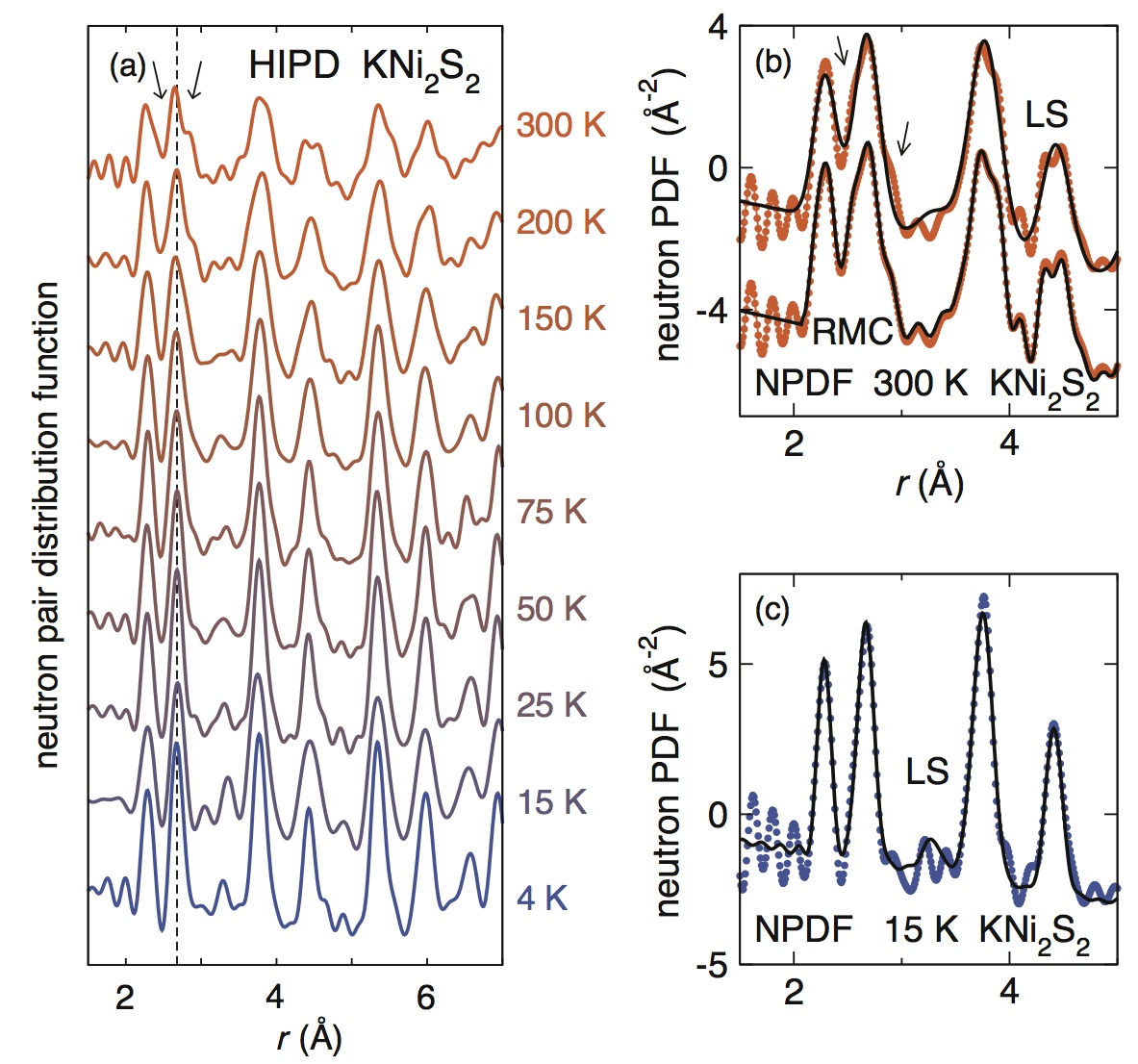}
\caption{ 
(a) Temperature dependence of the neutron pair distribution function derived from neutron total scattering data (HIPD, Lujan Center) of KNi$_2$S$_2$ shows the presence of multiple nearest-neighbor Ni--Ni distances observed at $T=300$~K that disappear on cooling (dashed line: ideal Ni--Ni separation). 
(b) Quantitative modeling of the PDF (NPDF, Lujan Center) measured at $T$ = 300~K reveals a poor least-squares fit of the $I4/mmm$ crystal structure determined by SXRD (LS) due to the presence of multiple Ni--Ni distances (arrows).  Reverse Monte Carlo (RMC) simulations accurately describe the data. 
(c) In contrast, the $T$ = 15~K PDF (NPDF) is quantitatively described by a least-squares fit of the average crystallographic structure with a single Ni--Ni distance.\label{fig:Tpdf} }
\end{figure}

Since there are no supercell reflections corresponding to a long range periodic order of the off-centering, pair distribution function (PDF) analysis of neutron total scattering data was used to probe the nature and spatial extent of the distortions.  Figure~\ref{fig:Tpdf}(a) shows the PDF analysis of total scattering data collected at $T=$ 300~K. Consistent with the off-centering, there are significant shoulders to the peak at $r \sim$ 2.68 \AA\ corresponding to modulations in nearest-neighbor Ni--Ni distances.  These displacements again only have a weak temperature-dependence at high temperature, but abruptly disappear below $T\sim$ 75~K,  concomitant with the observations from the SXRD analysis. Further, the ideal $I4/mmm$ crystal structure from the SXRD analysis provides an excellent fit to the $T$ = 15~K PDF [Figure~\ref{fig:Tpdf}(c), LS].  Ripples in the PDF with a period $\Delta r = 2\pi/ Q_\text{max} = 0.18$~\AA, amplified at low $r$, are artifacts the finite Fourier transformation used extract the PDF from the scattering data.\cite{Egami_Billinge} In contrast, while a split-site displacement of the Ni position describes the SXRD data at $T=300$~K,  it does not adequately describe the shoulders of the nearest-neighbor Ni--Ni correlation peak in the $T=300$~K PDF [Figure~\ref{fig:Tpdf}(b), LS and arrows].

\begin{figure}[t]
\centering
\includegraphics[width=3.5in]{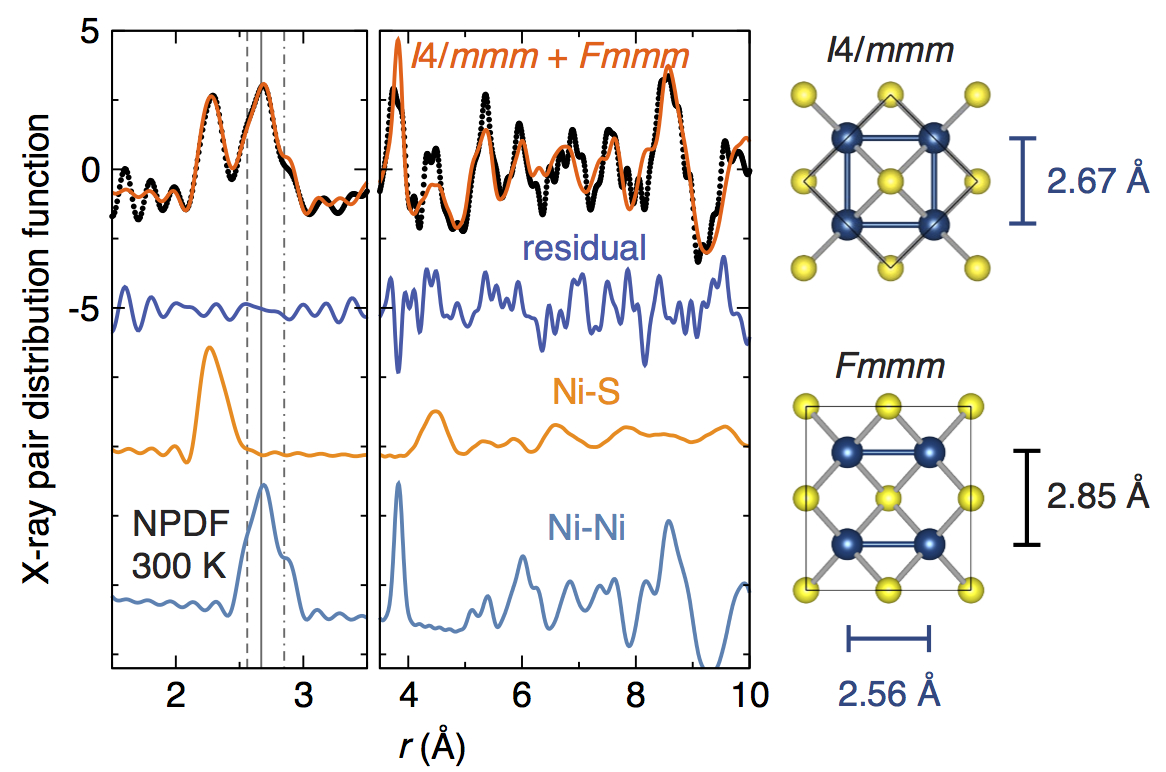}
\caption{Least-squares fit of a linear combination of tetragonal $I4/mmm$ and orthorhombic $Fmmm$ models of KNi$_2$S$_2$ to the $T$= 300K PDF allows extraction of the local Ni--S and Ni--Ni distances; however, the structural model fails to describe the data beyond $r>3.5$~\AA.\label{fig:LSpdf}}
\end{figure}

The $T=$ 300~K PDF requires at least three distinct Ni--Ni distances at $r\sim$ 2.57(1), 2.70(1), and 2.86(1) \AA\ (Figure~\ref{fig:LSpdf}), which are not provided by the split-site model used for Rietveld analysis of the SXRD data.  Deconvolution of the nearest-neighbor pair-wise correlations is provided by fitting a linear combination of  tetragonal ($I4/mmm$) and orthorhombic ($Fmmm$) phases to the PDF (Figure~\ref{fig:LSpdf}).  Each phase has split-site occupancy of the Ni atoms which are displaced along the $c$ axis.  All non-special internal coordinates and unit cell dimensions were allowed to refine along with the relative contribution of each phase.  The resulting relative contribution of each fraction of each phase is $f_{I4/mmm}$ =  48 at\% and $f_{Fmmm}$ = 52 at\% from least-squares refinement.  From the models, we extract three nominal Ni--Ni distances: $d_\text{Ni--Ni}$ = 2.56, 2.67, and 2.85 \AA, as illustrated at the right of Figure~\ref{fig:LSpdf}.  However, this structural model only describes the first coordination sphere of the Ni--S and Ni--Ni correlations; beyond $r>3.5$ \AA, even this multi-phase model fails to describe the PDF.

\begin{figure}[t]
\centering
\includegraphics[width=3in]{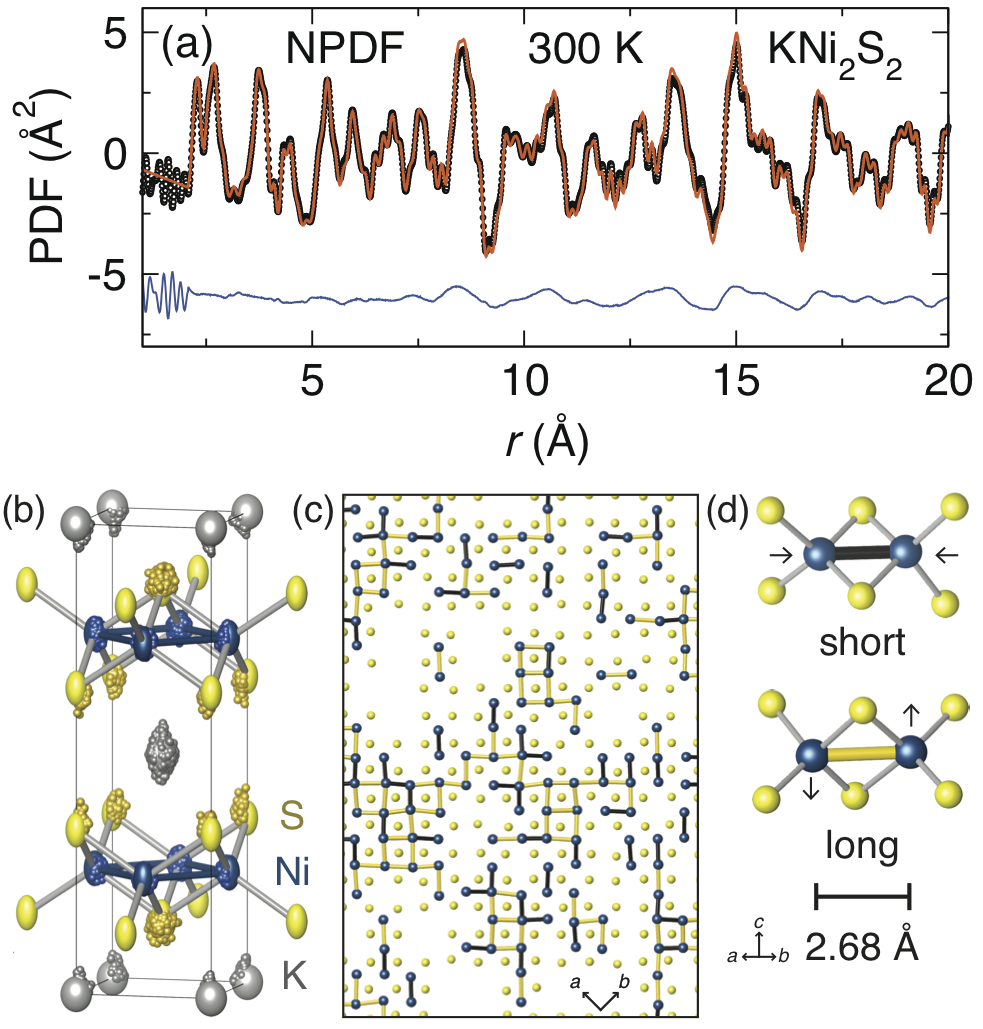}
\caption{
(a) Representative Reverse Monte Carlo simulation of the pair distribution function.  (b) Projection of all 24000 atoms from the RMC super-cell (small spheres) back onto the idealized crystallographic unit cell (large spheres) reveals that the ThCr$_2$Si$_2$-type connectivity of KNi$_2$S$_2$ is maintained after the simulation, as is the average $I4/mmm$ crystallographic symmetry.  (c) A representative [Ni$_2$S$_2$] layer extracted from a 24000 atom RMC supercell ($T=300$~K) illustrates the incoherent distribution of short (dark, $d_\text{Ni-Ni}<2.52$~\AA) and long bonds (light, $d_\text{Ni-Ni}>2.86$~\AA).  Intermediate length bonds are omitted for clarity. 
(d) Scaled atomistic representations of short and long Ni--Ni bonds extracted from the $T$ = 300~K RMC supercell.\label{fig:rmc}
 }
\end{figure}

Reverse Monte Carlo (RMC) simulations of the neutron pair distribution function and Bragg profile produce atomistic configurations that are compatible with the average crystallographic symmetry and extended pairwise correlations.   [Figure~\ref{fig:Tpdf}(b), RMC and Figure~\ref{fig:rmc}(a)].  Projection of all 24000 atoms from the large supercell back onto the crystallographic unit cell resembles the anisotropic atomic displacement parameters obtained from Rietveld analysis [Figure~\ref{fig:rmc}(b)].  Furthermore, the supercell does not reveal any locally ordered patterns [Figure~\ref{fig:rmc}(c)].  Instead, short and long Ni--Ni bonds appear randomly distributed throughout the lattice [Figure~\ref{fig:rmc}(d)].

\begin{figure}[t]
\centering
\includegraphics[width=3in]{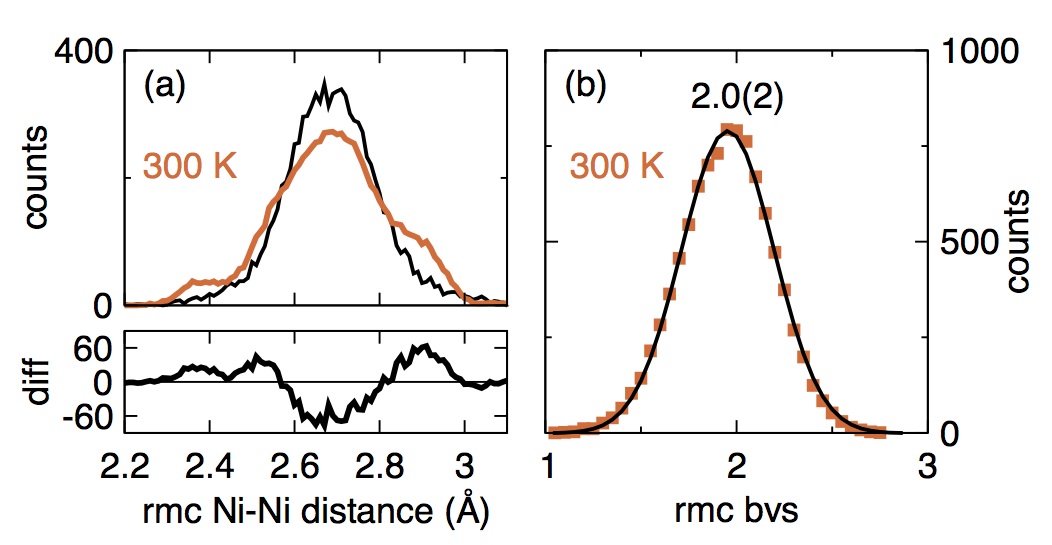}
\caption{(Color online)
(a) Histogram of Ni--Ni distances extracted from the RMC supercell.  The orange (light gray) line is the result from a simulation of the measured data.  The solid black line corresponds to a histogram of Ni--Ni distances simulated from anisotropic and harmonic displacements.  Below is their difference.  (b) The ensemble bond valence sums (BVS) indicate a single population of charged Ni species centered around a BVS = 2.0(2).  The solid line is a fit to a single Gaussian expression. \label{fig:bvs}
}
\end{figure}

Statistical analysis of the resulting RMC supercell yields an equivalent ensemble distribution of Ni--Ni distances, while also describing the extended pair-wise correlations.  The trimodal histogram of Ni--Ni distances [Figure~\ref{fig:bvs}(a)] appears with maxima centered around $r \sim$ 2.56, 2.67, and 2.85 \AA, consistent with the least-squares analysis [Figure~\ref{fig:LSpdf}].  To ensure this distribution is not the trivial result of harmonic but anisotropic atomic displacements, artificial  PDFs and Bragg profiles of KNi$_2$S$_2$ were generated using an ideal $I4/mmm$ unit cell with anisotropic thermal displacements, akin to Ref.~\onlinecite{Neilson_PRB_2012}.  These profiles were fit using  RMC simulations with the same starting supercell as used with the experimental data.  The histogram of Ni--Ni displacements from an anisotropic, but harmonically distorted structure has much more symmetric and singly distributed peak shape.  Subtraction of the control simulation from experimental distribution emphasizes and confirms the presence of three populations of bond lengths, as illustrated in Figure~\ref{fig:LSpdf}(a).  From the RMC supercell, we were able to extract an ensemble of the bond valence sums  (BVS) over all Ni--S distances contained within [NiS$_4$] tetrahedra.\cite{bondvalence}  The population is symmetrically distributed about a BVS = 2.0, with a full-width-at-half-maximum of 0.2, as opposed to a non-integer value or mixed-valence distribution.  

\begin{figure}[h]
\centering
\includegraphics[width=3in]{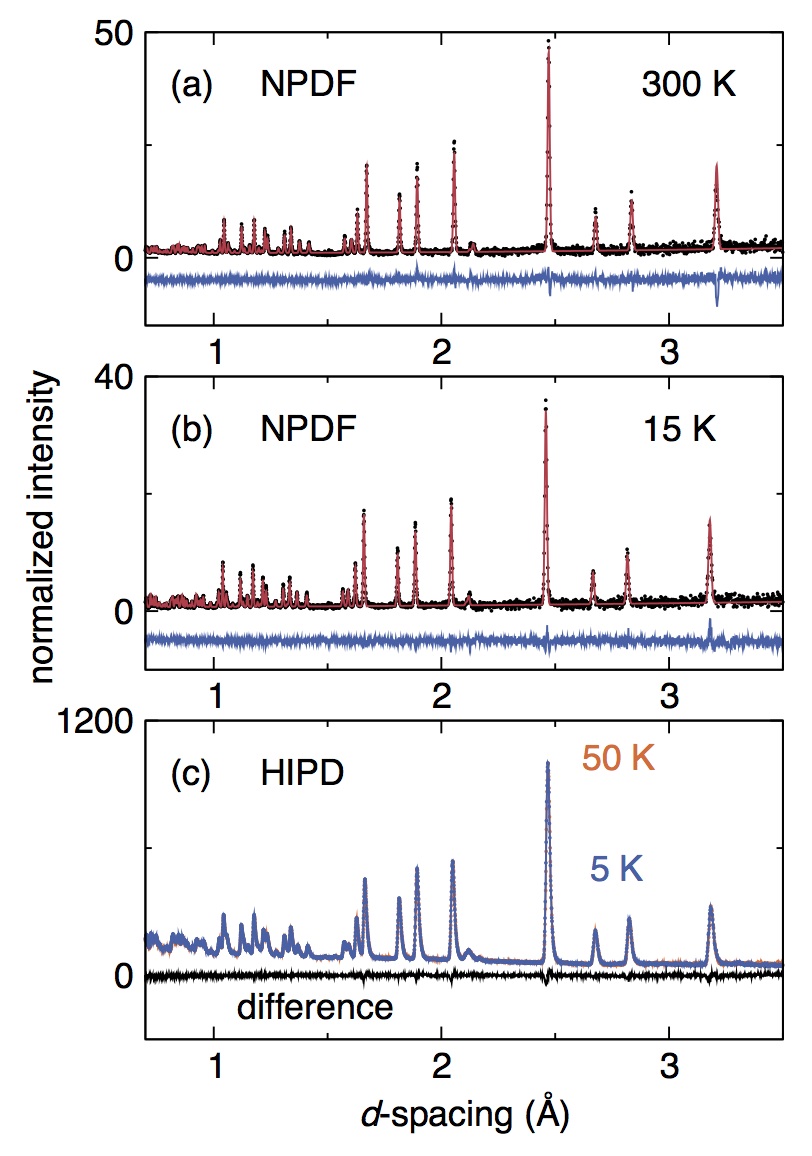}
\caption{
(Color online)
Neutron powder diffraction (NPD) data and from the NPDF instrument (90$^\circ$ bank) acquired at (a) $T$ = 300~K and (b) $T$ = 15~K and Rietveld analyses illustrating the absence of additional reflections, as would appear from long-range ordered magnetism and enlargement of the $I4/mmm$ nuclear unit cell (circles: data; line: fit; difference curve below).  (c) Direct subtraction of NPD data (black) collected at $T$ = 50~K (orange/light gray) and $T$ = 5~K (blue/dark gray) on the HIPD instrument to illustrate an absence of additional scattering from long-range ordered magnetism.  \label{fig:npd}}
\end{figure}

In these analyses, there is no evidence for long-range ordered magnetism.  The structural models used to describe the high-resolution synchrotron diffraction data provide excellent fits to the neutron powder diffraction data [Figure~\ref{fig:npd}(a,b)].  Direct subtraction of data sets collected at $T$ = 50~K and 5~K [Figure~\ref{fig:npd}(c)] does not reveal the appearance of any additional scattering, as would emerge from magnetic order.

\subsection{Physical Properties}

\begin{figure}[h]
\centering
\includegraphics[width=3.in]{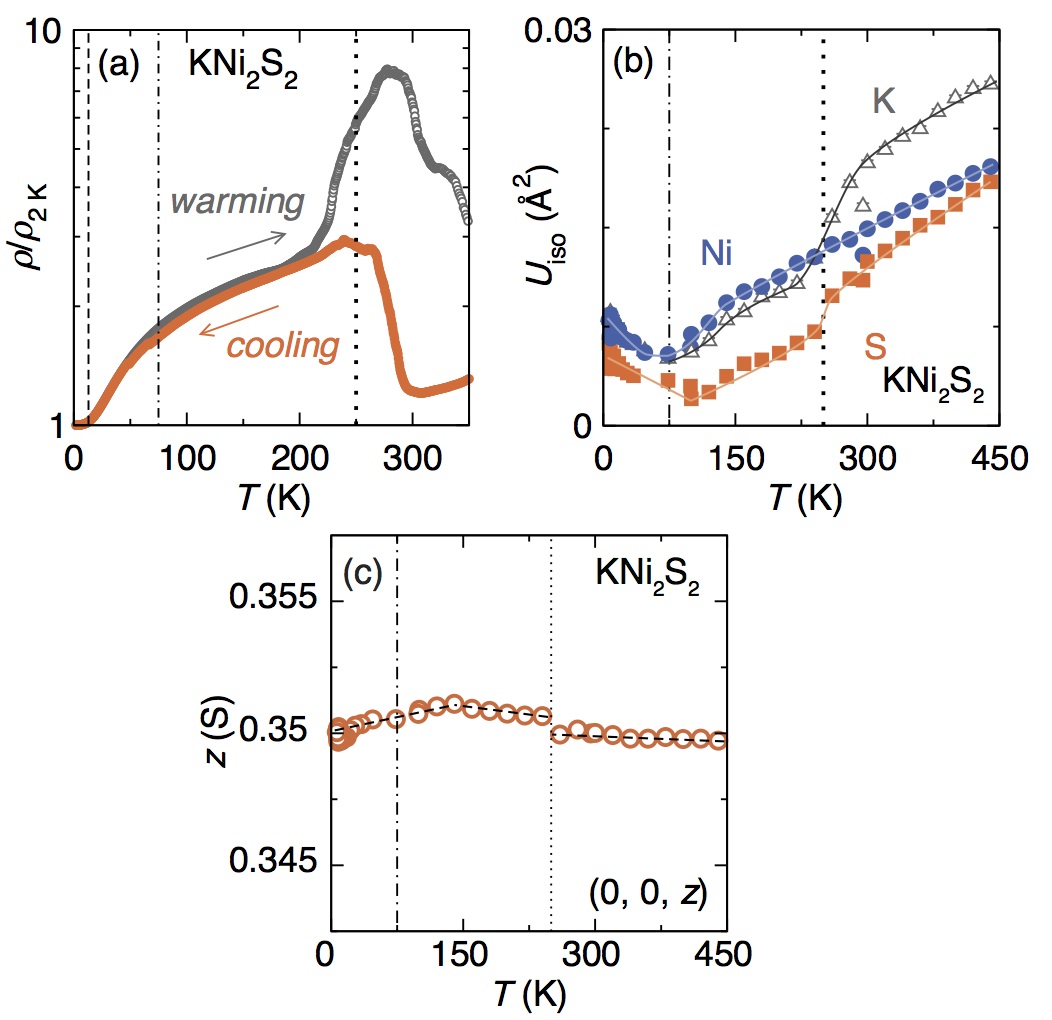}
\caption{
(a) Temperature dependence of electrical resistivity of KNi$_2$S$_2$, measured on a polycrystalline pellet, indicates metallic transport and a first-order transition near $T \sim$ 250~K. 
(b) The isotropic atomic displacement parameters for K, Ni, and S show transitions concomitant with the K$^+$ sublattice freezing at $T\sim$ 250~K (dotted line).  Colored lines are guides to the eye. (c) Rietveld analysis of the SXRD data reveals that the S position remains undisturbed, with only a minor transition observed at $T\sim$ 250~K.  \label{fig:freeze}}
\end{figure}

Resistivity measurements indicate metallic behavior at all temperatures, as previously reported [Figure~\ref{fig:freeze}(a)].\cite{Huan:1989p1663}  There is a discontinuity near $T\sim$ 250~K with hysteresis, as observed from first-order phase transitions.  This transition coincides with a significant change in the isotropic thermal displacement parameter of the K$^+$ sublattice as inferred from the SXRD data [Figure~\ref{fig:freeze}(b)]; however, neither the S position [Figure~\ref{fig:freeze}(c)] nor the Ni position are greatly disturbed [Figure~\ref{fig:11bmdistortion}(d)].

\begin{figure}[t]
 \centering
\includegraphics[width=2.5in]{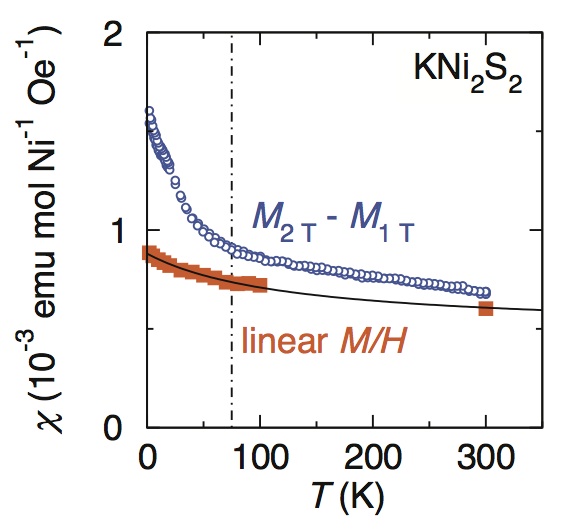}
\caption{
The linear magnetic susceptibility (closed squares; thin line as a guide to the eye), as determined by the high-field slope of  isothermal magnetization, is nearly temperature-independent.  The magnetic susceptibility determined by measurements at constant fields (open circles) reveals a gradual upturn at $T\sim$ 75~K.
\label{fig:chi}
}
\end{figure}

\begin{figure}[t]
\centering
\includegraphics[width=3in]{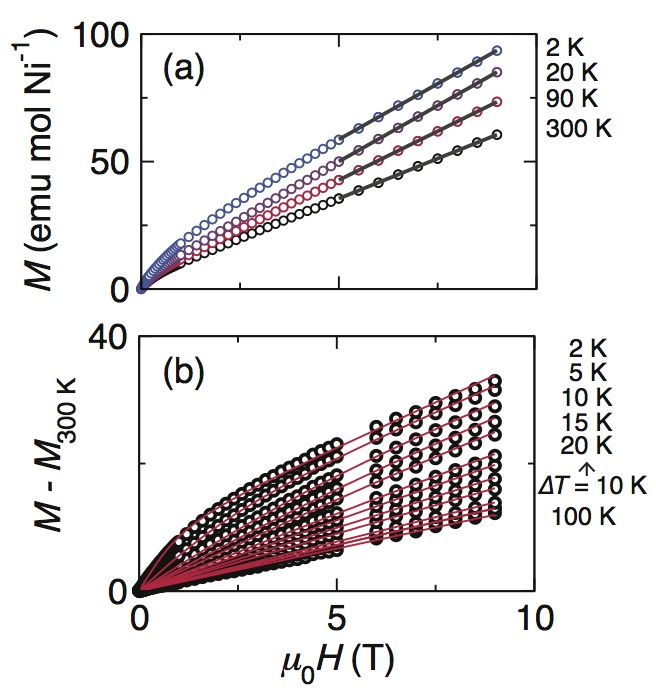}
\caption{
(a) Isothermal field-dependent magnetization curves at various temperatures (open circles) illustrate the linear contribution at high magnetic field  (solid lines) and increased curvature at low temperatures.  
(b) Global fit (thin lines) of a Brillouin function and linear susceptibility term (Eqn.~\ref{eqn:brillfit}) to the $M-M_\text{300~K}$ data (open circles).  
 \label{fig:isomag}}
\end{figure}

Measurement of the linear magnetic susceptibility reveals only a weak temperature dependence (Figure~\ref{fig:chi}).   The magnetization was measured in two ways: isothermally and at constant field.  The constant-field magnetic susceptibility ($\chi$) was approximated by, $\chi \approx \Delta M/\Delta H = [M_{\text{2T}}-M_{\text{1T}}]/[1\text{T}]$, in order to subtract trace ferromagnetic Ni impurities that give a subtle curvature to the $T=300$~K magnetization, as shown in Figure~\ref{fig:isomag}(a)  (impurity concentration $<1$\%, undetectable by SXRD).  The constant-field susceptibility exhibits a gradual upturn below $T<75$~K.   To test if this upturn results from a contribution of localized moments following a Brillouin function,  many isothermal field-dependent magnetization measurements were measured for $T<$ 100~K.  The curvature of the isothermal magnetization, $M(H)$, pictured in Figure~\ref{fig:isomag}(a) and (b), is not well described solely by a Brillouin function: the magnetization for $\mu_0H >$ 5~T is linear, even down to $T=2$~K.  

To extract the trace ferromagnetic impurity from the isothermal magnetization curves, the $T=$ 300~K magnetization data were subtracted from the $T\le100$~K data.  To account for the remaining curvature of the data, we performed a global fit of all of the magnetization data to,
\begin{equation}
M  - M_\text{300~K} = n g J B_J(x) + [\chi(T)-\chi_\text{300K}]\ H  \label{eqn:brillfit}
\end{equation}
where $\chi(T)$ is the temperature-dependent linear susceptibility, $\chi_\text{300K}$ is the linear slope of $M$($H$, 300~K), $n$ is number of localized paramagnetic spins per mol Ni, $g$ is the gyromagnetic ratio, $J$ is the total angular momentum, $x =(\mu_BH)/(k_BT)$, $H$ is the applied magnetic field,  and 
\begin{equation}
B_J(x) = \frac{2J+1}{2J}\coth\left(\frac{(2J+1)x}{2J}\right) - \frac{1}{2J}\coth \left(\frac{x}{2J}\right).\label{eqn:brill}  
\end{equation}
From the global fit with $g=2$, we extracted $J$ = 2.0(1) and $n = 3.2(1)\times 10^{-4}$  impurity spins per mol Ni.  The values of $\chi(T)$ are shown in Figure~\ref{fig:chi} as linear $M/H$.  We obtain  equivalent values of $\chi$ by simply extracting the slope of the linear portions of the magnetization curves [$\chi= \Delta M / \Delta H$; Figure~\ref{fig:isomag}(a)].  Because the  values of $\chi$ are very small ($\sim10^{-3}$ emu mol Ni$^{-1}$ Oe$^{-1}$), trace impurities (Ni$^{2+}$, $J \sim$ 1 to 4) have a significant effect on the observed magnetization.

The linear contributions to the isothermal magnetization follow a weak temperature dependence; fitting these values [closed squares, Figure~\ref{fig:chi}] to the Curie-Weiss equation, $\chi = C/(T-\Theta) + \chi_0$, allows us to calculate a lower bound to temperature-independent contribution of the magnetic susceptibility, $\chi_0 > 4.9(5)\times10^{-4}$ emu mol Ni$^{-1}$ Oe$^{-1}$.

\begin{figure}[t]
 \centering
 \includegraphics[width=3in]{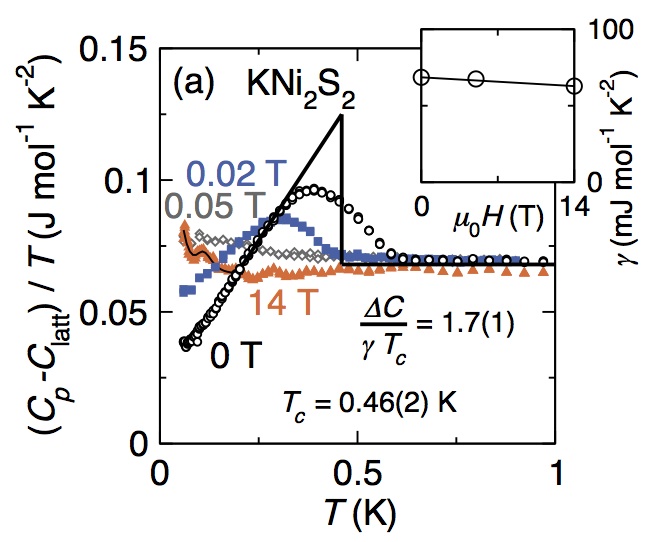}
\caption{
The electronic specific heat ($C_p-C_\text{latt}$) reveals a bulk superconducting transition at $T_c$ = 0.46(2)~K (thick solid line: equal entropy construction).  Magnetic fields suppress the transition, and by $\mu_0H$= 14 T  a sharp upturn at $T<$~0.2~K  is present and described by Schottky anomalies corresponding to impurity ($\sim10^{-6}$) and nuclear spins (thin solid line).  Inset: applied magnetic fields have a weak influence on $\gamma$.  \label{fig:schc}
}
\end{figure}

The low-temperature specific heat data (Figure~\ref{fig:schc}) reveals a $\lambda$-type anomaly consistent with bulk superconductivity at $T_c$ = 0.46(2)~K.  For $1.8 < T<20$~K, the total specific heat was modeled as, $C = \gamma T + \beta_3 T^3 + \beta_5 T^5$, to extract the electronic contribution to the specific heat described by the Sommerfield coefficient, $\gamma$ = 68(1) mJ mol$^{-1}$ K$^{-2}$ (Table~\ref{tab:hc}).   The specific heat jump at the transition is $\Delta C_e / \gamma T_c$ = 1.7.  Small external magnetic fields suppress the superconducting transition, with $H_{c2}$(0K)$\sim$0.04(1)~T, obtained by fitting the observed field dependence of $T_c$ to a two-fluid model.    The Sommerfield coefficient is only weakly dependent on an applied magnetic field [Figure~\ref{fig:schc}, inset].  The small upturn in the heat-capacity at $\mu_0H$ = 14 T for $T<$ 0.2~K is well described by Schottky anomalies for nuclear and impurity spins.

\begin{figure}[t]
 \centering
 \includegraphics[width=3.in]{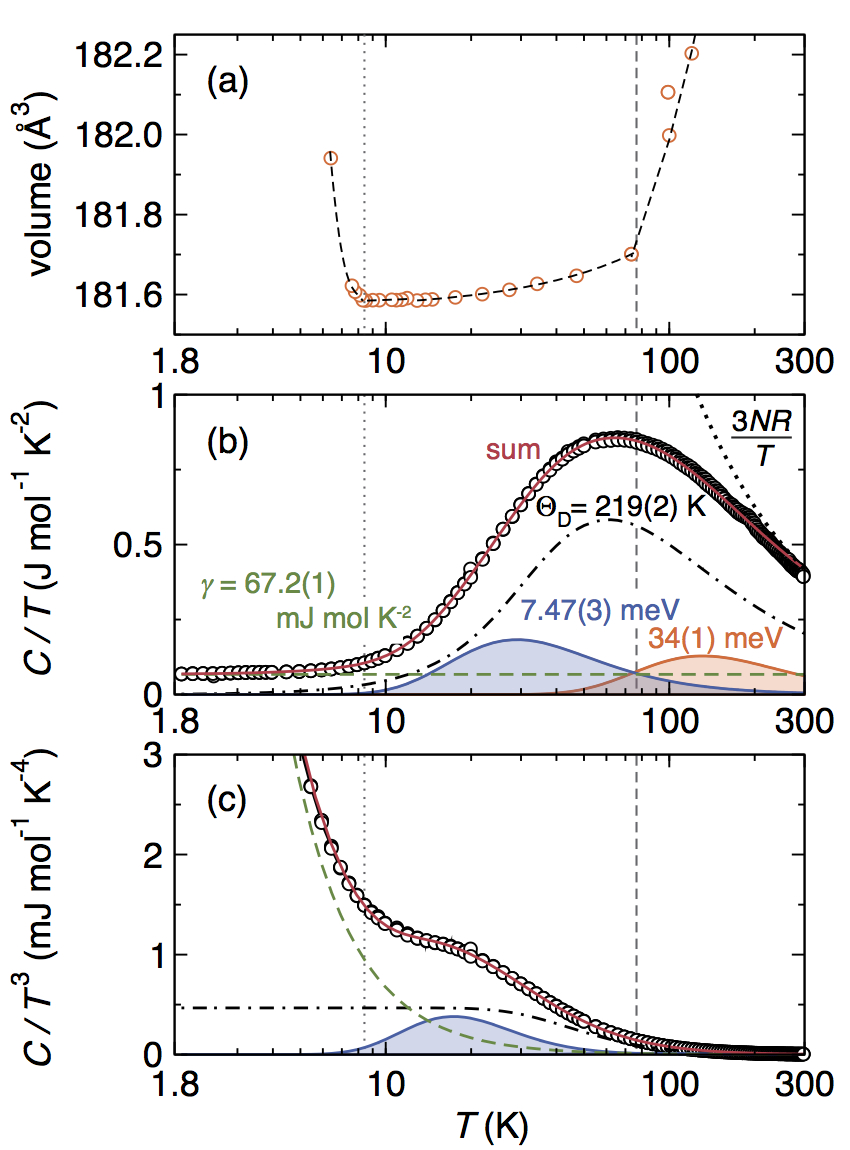}
\caption{ 
(a) The temperature dependence of the unit cell volume of KNi$_2$S$_2$ shows two anomalies  below $T<13$~K, with an overall negative coefficient of thermal expansion below $T<8.9$~K.   (dashed lines guide the eye, vertical lines denote transitions).  
(b) The excess entropy  below $T\sim$ 10 K is only well described when an enhanced-mass $\gamma T$ electronic contribution is included.  The Dulong-Petit limit ($3NR/T$; dotted line) is achieved near $T=300$~K.  
(c) The specific heat normalized by $T^3$ indicates additional thermodynamic degrees of freedom, well described by the Einstein expression for localized vibrational modes, as described in the text ($\gamma T$ contribution: dashed line; Debye lattice: black dot-dashed line; Einstein modes: shaded peaks). 
\label{fig:einstein}
}
\end{figure}

Normalizing the specific heat measured to higher temperatures by $T^3$ [Figure~\ref{fig:einstein}] reveals non-dispersive phonon contributions to the lattice heat capacity.  Dispersive phonons should plateau when plotted as $C/T^{3}$ with decreasing temperature;\cite{PhysRevB.79.224111,PhysRevLett.80.4903} meanwhile the electronic contribution rises sharply  ($C_e/T^3= \gamma/T^2$).  Therefore, the high-temperature specific heat was fit to combination of a Debye lattice model and several Einstein modes describing non-disperseive, localized lattice vibrations.   The total heat capacity was described as:
\begin{align}
\begin{split}
C_v = & \gamma T +   9Rs\left(\frac{T}{\Theta_{D}}\right)\int^{x_{D}}_0 \frac{x^4e^xdx} {(e^x-1)^2} + \\
&\sum_{i=1}^2 p_i R \frac{(\hbar \omega_i/k_BT)^2 e^{\hbar \omega_i/k_BT}}{(e^{\hbar \omega_i/k_BT}-1)^2}
\label{eqn:debye}
\end{split}
\end{align}
where $R$ is the gas constant, $\Theta_{D}$ is the Debye temperature, $x_D = \Theta_D / T$, $s$ is the number of Debye oscillators, $\hbar \omega_i$ is the energy of the $i^\text{th}$ dispersionless mode, and $p_i$ is the number of dispersionless oscillators.   By  simultaneously fitting the measured heat capacity to Eqn.~\ref{eqn:debye} as $C/T$ (in J mol$^{-1}$ K$^{-2}$) and as $C/T^2$ (in J mol$^{-1}$ K$^{-3}$), it ensures a proper weighting of the high-temperature curvature of the Debye expression and the low-temperature curvature of the Einstein expressions and electronic contributions, respectively.  The model provides an excellent fit to the experimental heat capacity data; the fit parameters are tabulated in Table~\ref{tab:hc}.  Two localized (dispersionless) vibrational modes, each described by an Einstein expression with energies, $\hbar \omega$ = 7.47(3)~meV [87(4)~K] and 34(1)~meV [394(12)~K] are found in addition to contributions from Debye expression for dispersive phonon and the conduction electrons. Fits with only a single Einstein mode do not provide description of the data, but we cannot rule out the possibility that there are more than two Einstein modes on the basis of these data.

\begin{table*}[t]
\centering
\caption{
 Specific heat fitting parameters}. \label{tab:hc}
 \begin{tabular*}{\textwidth}{@{\extracolsep{\fill}}c||cc||cc|cc}
    \hline
    \hline
 Electronic term  & \multicolumn{2}{c||}{Debye terms}& \multicolumn{4}{c}{Einstein terms} \\
   $\gamma$ (mJ mol$^{-1}$ K$^{-2}$) & $\Theta_D$ (K) & $s$ (osc.)  & $\hbar \omega_1$ (meV) & $p_1$ (osc.) & $\hbar \omega_2$ (meV) & $p_2$ (osc.)  \\
  \hline
  67.6(1)    &  219(2)  & 2.52(5)   &  34(1)  & 1.50(5)  & 7.47(3)  & 0.47(1) \\
    \hline
    \hline
   \end{tabular*}
\end{table*}

\section{Discussion}

The subtle, but rich structural transitions observed in KNi$_2$S$_2$ are indicative  of strongly-correlated electronic physics.  The fact that the Ni displacements disappear on cooling is a highly unusual observation, an indicator of an increase in local symmetry on cooling.  To our knowledge, the only materials in which this has been observed are the colossal magnetoresistive perovskite manganites,\cite{Louca:1999wh, Rodriguez:2005gt,Bozin:2007uz}  PbRuO$_3$,\cite{Kimber:2009wc} the binary lead chalcogenides,\cite{Bozin:2010iw} and the analogous compound, KNi$_2$Se$_2$.\cite{Neilson_PRB_2012}

While many materials exhibiting a negative coefficient of thermal expansion (NTE) are known, in most examples, such as ReO$_3$,\cite{Rodriguez:2009tf} a NTE typically arises from the connectivity of rigid polyatomic units that become less flexible on cooling or from the increased amplitude of rigid unit modes on heating that pulls in the structure, as with the analogy to a guitar string\cite{Giddy:br0023}.  Such behavior is not expected in KNi$_2$S$_2$ due to the constrained connectivity of edge-sharing [NiS$_4$] tetrahedra within the layers.  Instead, the negative thermal expansion of KNi$_2$S$_2$ may reflect the breaking of directional bonds and delocalization of charge,  as illustrated by c\ae sium, where the addition of hydrogen results in a \emph{decrease} in volume per formula unit from Cs (111.7  \AA$^3$/Cs) to CsH (64.8 \AA$^3$/CsH).\cite{Cs,CsH}   This comparison suggests that we observe the formation of a more delocalized electronic state in KNi$_2$S$_2$ at the lowest temperatures and analogous to the driving force behind anomalous thermal expansion in heavy fermion materials (\emph{e.g.}, URu$_2$Si$_2$).\cite{RevModPhys.56.755,Fetisov1985403}  These observations may also be related to intermediate valence compounds, YbCuAl\cite{Mattens1980982} or CeAl$_3$,\cite{PhysRevLett.35.1779} where a more spatially extended valence state is favored at lower temperatures.  

However, here, such an effect requires an involvement of direct Ni--Ni bonding, as the bond valence sum analysis of the Ni--S bonding does not show any evidence for charge disproportionation or Jahn-Teller distortions in KNi$_2$S$_2$.\cite{Shoemaker:2009wh,Shoemaker:2010ia}  Instead, our results and the absence of charge order are consistent with the off-centering displacements arising from chage density wave (CDW) fluctuations within a manifold comprised predominately of Ni $d_{x^2-y^2}$ orbitals, as proposed for KNi$_2$Se$_2$.\cite{Neilson_PRB_2012}   These CDW fluctuations are spatially related to not only the long-range ordered CDWs observed in KCu$_2$Se$_2$ and SrRh$_2$As$_2$,\cite{PhysRevB.67.134105, PhysRevB.85.014109}  but also the tetragonal-to-orthorhombic structural distortions observed in BaFe$_2$As$_2$ \cite{Rotter:2008p5432} and Fe$_{1.01}$Se.\cite{PhysRevLett.103.057002}

Quantitative analysis of the specific heat at higher temperatures reveals insight into the dynamic nature of the structural anomalies as charge density wave fluctuations.     On cooling, most of the entropy of the 34(1) meV mode is released at a similar temperature scale to the disappearance of the CDW fluctuations; the entropy of the 7.47(3) meV mode is released at the same temperature scale as the observation of negative thermal expansion behavior $T<8.9$~K (Figure~\ref{fig:einstein}).  Analysis of the heat capacity is consistent with both the loss of a CDW around T$\sim$75 K and the emergence of heavy electron physics at low temperature, suggesting that the two phenomena are interrelated.  

While a significant increase in carrier mobility concomitant with the depopulation of Ni displacements was observed in KNi$_2$Se$_2$,\cite{Neilson_PRB_2012} the transport properties of KNi$_2$S$_2$ appear to be dominated by the freezing of the K$^+$ sublattice [Figure~\ref{fig:freeze}(a)], as inferred from the significant decrease in the K $U_{iso}$ obtained from the SXRD data [Figure~\ref{fig:freeze}(b)].  Such transitions are usually first-order due to an associated latent heat, as per liquid-solid transitions,\cite{PhysRevLett.32.596} and potassium has been shown to be readily mobile in the related compound, KNi$_2$Se$_2$ at room-temperature.\cite{Neilson_JACS_2012}   This freezing process likely introduces  many electronic scattering centers and prevents an accurate measurement of the intrinsic resistivity.  

We observe two significant thermodynamic signatures of many-body physics of KNi$_2$S$_2$  in addition to the negative thermal expansion: superconductivity and an enhanced electronic band mass.   The low-temperature specific heat data [Figure~\ref{fig:schc}] reveals a $\lambda$-type anomaly consistent with bulk superconductivity at $T_c$ = 0.46(2)~K.  
Such a small $H_{c2}$ (compared to $T_c$) is indicative of an enhanced mass of conduction elections, between $m_{H_{c2}}^*$ = 40$m_e$ to 80$m_e$.  This estimate of the degree of electronic mass enhancement depends on the assumptions about the size and shape of the Fermi surface.  Using the extrapolated $T=0$~K upper critical field, $H_{c2}$ = 0.04(1) T, the average zero-temperature coherence length of the superconducting state is $\xi(0) = [\phi_0/(2\pi H_{c2})]^{1/2}=$ 90(10)~nm.\cite{morosan-prb2012}  The Fermi velocity is estimated from $T_c$ and $\xi$ using the relation $v_F=(k_B T_c \xi)/(0.18\hbar)= 3(1)\times 10^4$ ~m/s. Then, assuming a spherical Fermi surface, the Fermi wavevector is estimated from the carrier density, $n$ = carriers per unit cell volume, using $k_F=(3n\pi^2)^{1/3}$.  If there are 1.5 carriers per Ni, then $k_F$ = 0.99(1) \AA$^{-1}$, while if all 33 valence electrons per formula unit contribute, then $k_F$ = 2.2(1) \AA$^{-1}$. The resulting mass enhancement is then calculated from the relation $m^*/m=\hbar k_F /v_F$, and ranges from $m_{H_{c2}}^*/m$ = 40 to 80. 

The specific heat in the normal state above $T_c$ also points to an enhanced band mass: the low-temperature Sommerfield coefficient, $\gamma$ = 68(1) mJ mol$^{-1}$ K$^{-2}$, represents a significant electronic mass enhancement, between $m^*$ = 11$m_e$ to 24$m_e$.  To estimate the electronic band mass enhancement, we assume a spherical Fermi surface with either 33 valence electrons per formula unit, or 3 valence electrons per formula unit (Ni$^{1.5+}$).  The carrier density, $n$, is given by the number of carriers, $N$, per cell volume, $V$, to give $n = N/V$, which is used  in the calculation of the Fermi wavevector, $k_F = (3\pi^2n)^{1/3}$, in order to estimate the densities of states at the Fermi energy, $g(E_F)$.  The Sommerfield coefficient from the spherical Fermi surface is given by $\gamma_e = \pi^2 k_B^2    g(E_F)/3$.  The electronic band-mass enhancement is then estimated from $m^*/m_e = \gamma_\text{measured} / \gamma_e$.  

This electronic mass enhancement, comparable to that extracted from the upper critical field, represents a much larger mass enhancement than is generally observed in metallic correlated solids \cite{PhysRevB.79.245114,Mcqueen:2009p7274} and is comparable the mass enhancement observed in the archetypical heavy fermion compound, URu$_2$Si$_2$.\cite{PhysRevLett.56.185}  Unlike prototypical heavy fermion materials,\cite{PhysRevB.41.9352,JPSJ.70.2248,PhysRevLett.78.3729} however, the Sommerfield coefficient is only weakly dependent on an applied magnetic field [Figure~\ref{fig:schc}, inset], suggesting at most a minor role for magnetism or magnetic fluctuations in producing the heavy mass state. This is additionally supported by the lack of localized magnetism observed in magnetic susceptibility measurements (Figure~\ref{fig:chi} and Figure~\ref{fig:isomag}) and high-flux neutron powder diffraction experiments (Figure~\ref{fig:npd}).

\begin{figure}[t!]
 \centering
 \includegraphics[width=3in]{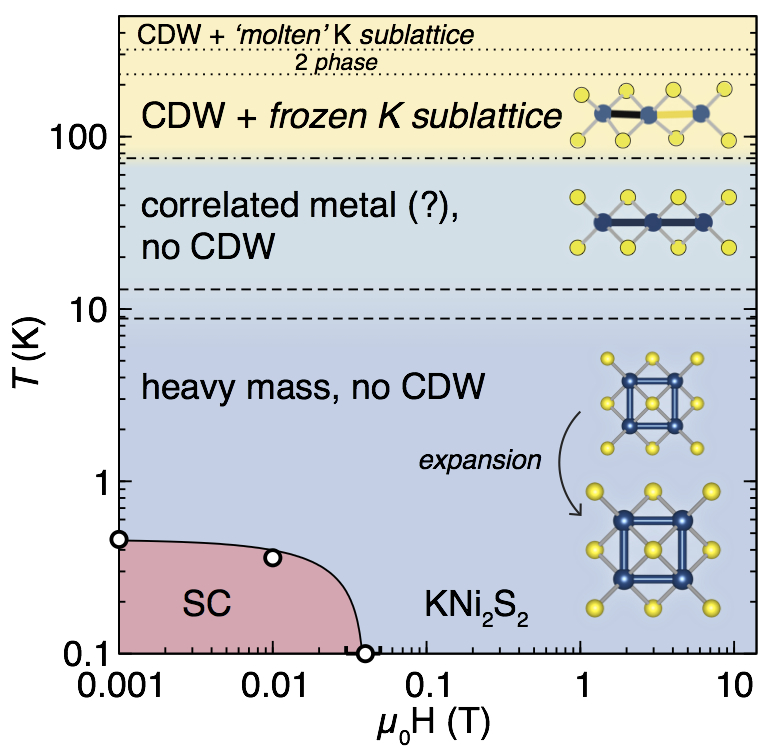}
\caption{
The temperature/magnetic-field phase diagram displays the complex structural and electronic states observed in KNi$_2$S$_2$, including superconducting (SC), heavy mass, intermediate but non-charge density wave (CDW), Ni--Ni  charge density wave, and high K$^+$ mobility phases.  The K$^+$ sublattice melting transition is a first-order transition at $T=$ 250~K with an associated two-phase region.  The other transitions appear to be second-order, but the data are ambiguous regarding their thermodynamic nature.  
\label{fig:phase}
}
\end{figure}

\section{Conclusions}

In short, our data show that KNi$_2$S$_2$ exhibits a rich and unusual electronic and structural phase diagram below $T\sim$ 440~K, as summarized in Figure~\ref{fig:phase}.  Near room temperature, the K$^+$ sublattice is mobile and exhibits what appears to be a freezing transition near $T\sim$ 250~K.  Above $T\sim$ 75 K, there are aperiodic, incoherent CDW fluctuations corresponding to displacements of the Ni--Ni sublattice, concomitant with complete release of the entropy from a localized vibrational mode at $\hbar \omega$ = 34(1) meV.  Below $T\sim75$~K, we observe an intermediate state, which can be described as a correlated metallic state in which there is no CDW nor significant electronic mass enhancement.  Upon further cooling, KNi$_2$S$_2$ displays a significantly enhanced electronic mass, 11$<m^*/m_e<$24, with a concomitant lattice expansion [negative coefficient of thermal expansion] and the entropy release of a dispersion less vibrational moe of 7.47(3) meV.  Below $T_c$ = 0.46(2)~K there is a superconducting transition that is suppressed by a $H_{c2}$(0K) = 0.04(1)~T.  Surprisingly, all of these strongly correlated behaviors occur in the absence of localized magnetism. Instead, our results suggest that the origin of heavy electron behavior in KNi$_2$S$_2$ lies in the hybridization of nearly localized and bonded states with conduction electrons.\cite{Neilson_PRB_2012,MurrayTesanovic} It will be interesting to establish how proximity to charge order can drive strongly correlated physics in this and related materials families.


\section*{Acknowledgements}

J.R.N. and T.M.M thank C.~Broholm, Z.~Tesanovic,  J.~Murray, and O.~Tchernyshyov for helpful discussions.  This research is principally supported by the U.S. DoE, Office of Basic Energy Sciences (BES), Division of Materials Sciences and Engineering under Award DE-FG02-08ER46544.  The dilution refrigerator was funded by the National Science Foundation Major Research Instrumentation Program, Grant \#NSF DMR-0821005. This work benefited from the use of HIPD and NPDF at the Lujan Center at Los Alamos Neutron Science Center, funded by DoE BES. Los Alamos National Laboratory is operated by Los Alamos National Security LLC under DoE Contract DE-AC52-06NA25396.  The upgrade of NPDF was funded by the National Science Foundation through grant DMR 00-76488.  This research has benefited from the use of beamline 11-BM at the Advanced Photon Source at Argonne National Laboratory, supported by the U. S. Department of Energy, Office of Science, Office of Basic Energy Sciences, under Contract No. DE-AC02-06CH11357.   

\bibliography{../../../../../../Neilson_References}

\end{document}